\documentstyle[prl,aps]{revtex}
\begin{document}
\draft
\wideabs{
\title{LOCAL MAGNETISM OF ISOLATED Mo ATOMS AT SUBSTITUTIONAL AND
INTERSTITIAL SITES IN  Yb METAL: EXPERIMENT AND THEORY}

\author{ A.A.  Tulapurkar$^1$, S.N. Mishra$^1$, R.G. Pillay$^1$,  H.G. 
Salunke$^2$, G.P. Das$^2$ and  S. Cottenier$^3$ } 
\address{1 Department of Nuclear and Atomic Physics, T.I.F.R, Homi Bhabha
Road, Mumbai-400005, INDIA.} 
\address {2 TP $\&$ PE Division, Bhabha Atomic Research Center, Trombay,
Mumbai-400085, INDIA.}
\address{3  Instituut voor Kern- en Stralingsfysica, KU Leuven,
Celestijnenlaan 200D, B-3001 Leuven, Belgium.} 
\maketitle
\begin{abstract}
Using TDPAD experiment and local spin density calculations, we have
observed large 4d moments on isolated Mo atoms at substitutional and
octahedral interstitial lattice sites in Yb metal, showing Curie-Weiss
local susceptibility and Korringa like spin relaxation rate.  As a
surprising feature, despite strong hybridization with the Yb neighbours,
interstitial Mo atoms show high moment stability with small Kondo
temperature. While, magnetism of Mo, at substitutional site is consistent
with Kondo type antiferromagnetic d-sp exchange interaction, we suggest
that moment stability at the interstitial site is strongly influenced by
ferromagnetic polarization of Yb-4f5d band electrons.

\end{abstract}
} 
The formation and stability of local magnetic moments on d-impurities in
metallic hosts has been a topic of intense experimental and theoretical
investigations over the past several years.  While extensive studies have
been made for several 3d impurities [1,2], much less information is
available on the magnetism of 4d atoms in metals.  In general, the
d-electrons in 4d metals are regarded as being itinerant and do not show
tendency towards local moment formation.  Recently, applying time
differential perturbed angular distribution/correlation (TDPAD/TDPAC)
methods, strong magnetic behaviour has been observed for some
substitutional 4d impurities in few metals and alloys [2,3].  The results
revealed that, similar to the behaviour observed for 3d impurities,
magnetism of 4d ions in metals strongly depends on the type of conduction
electrons in the host. Furthermore, it has been argued that moment
stability measured by the Kondo temperature, T$_{K}$ is strongly
influenced by induced spin polarization of host band electrons.

In view of these aspects of magnetism for substitutional 3d and 4d
impurities in metals, one can ask : can local moment occur on 3d/4d atoms
at interstitial lattice sites also?  If so, how stable are the moments?
What is the magnitude and sign of host spin polarization and how does it
influence the impurity magnetism? Intuitively, due to reduced interatomic
distances and the consequent increase in the hybridization strength, one
would expect suppressed magnetism for d atoms at interstitial lattice
sites. However, large 3d moment has recently been observed for Fe at
interstitial sites in fcc Yb metal [4]. Hitherto, magnetism of 4d atoms at
interstitial lattice sites in a metal has not been investigated.  In this
letter we report experimental and theoretical studies on the magnetism of
isolated Mo impurity atoms at substitutional and interstitial lattice
sites in Yb metal.  The results, obtained from TDPAD measurements and
local spin density (LSD) calculations show that Mo atoms occupying
substitutional and octahedral interstitial sites posses large stable 4d
moments with rather small T$_{K}$ values. We show that lattice site
dependent magnetism of Mo, especially the moment stability, is strongly
influenced by the sign and strength of host conduction electron spin
polarization.

TDPAD experiments were carried out at the TIFR/BARC Pelletron accelerator
facility, Mumbai. We used the 8$^+$ isomer in $^{94}$Mo (T$_{1/2}$ = 98ns,
g$_{N}$ = 1.31) as a nuclear probe for the detection of magnetic response.
Heavy ion reaction $^{82}$Se($^{16}$O,4n)$^{94}$Mo with pulsed $^{16}$O
beam of energy 62 MeV was used to produce and recoil implant the probes in
Yb hosts. The estimated concentration of Mo in Yb was less than 1 ppm.
Spin rotation spectra, R(t) were recorded in the temperature range 10 to
300 K by applying magnetic field of 2 T.  Further details on the
experimental method can be found in Ref.  [5].  Measurements were
performed in fcc and hcp Yb.  Following Ref.[6], fcc Yb could be obtained
by rolling a piece of pure (99.9$\%$) metal to a thickness of about 5 to
10 mg/cm$^{2}$. To get the hcp phase, a thin disk of Yb metal was annealed
at 400$^\circ$C for 12 hours followed by repeated cycles of dipping in
liquid nitrogen and warming to 300 K. The crystal structures of the
samples were verified by X-ray diffraction measurements.

Figure 1 shows some examples of spin rotation spectra, R(t) for $^{94}$Mo
in fcc and hcp Yb along with their Fourier transforms. The spectra show
superposition of three frequency components arising from Mo atoms at
different lattice sites. They were fitted by the function [5]: R(t)
=$\sum_{i} A_{i}\exp(-t/\tau_{N_{i}}) sin[2(\omega_{L_{i}}t -
\theta)]$ to extract the amplitude A, Larmor frequency $\omega_L$ and the
nuclear spin relaxation time $\tau_N$ of each component. Fig.2 shows the
local susceptibilities, $\chi_{loc}(T) = \beta(T)$-1 of Mo deduced from
the relation $\omega_L(T) =
\hbar^{-1}g_N\mu_NB_{ext}\times\beta(T)$. It can seen that
$\beta$ for to two of the components in fcc as well as hcp Yb, strongly
vary with temperature which could be fitted to a Curie-Weiss law:
$\beta(T)-1 = C/(T+T_{K})$.  The Curie constants C for the two components,
in both phases of Yb, were found to be +15(2) K and -32(2) K.
Furthermore, the corresponding nuclear relaxation times $\tau_{N}$, shown
in Fig. 3, exhibit Korringa like behaviour ($\tau_{N} \propto$T).  Both
these features reflect strong local magnetism of Mo.  From the amplitudes
(A$_{i}$), it turns out that nearly 50$\%$ of the implanted Mo atoms in
fcc as well as hcp Yb show positive Curie constant while a fraction of
$\sim 25\%$ exhibit negative C value. The remaining $\sim 25\%$ show
temperature independent $\chi_{loc}$ with $\beta(T) = 1.00\pm0.02$.
Tentatively, we assign the components characterized by $\beta(T) <$1 and
$\beta(T) >$1, respectively to substitutional and octahedral interstitial
site. The third fraction ($\beta(T)$ = 1) presumably corresponds to a
tetrahedral interstitial site. Since fcc and hcp Yb has identical near
neighbour environments, for our later discussion of Mo magnetism, we
mainly consider the components ascribed to substitutional and octahedral
interstitial sites in fcc Yb.

To get theoretical understanding on the lattice site dependent magnetism
of Mo in Yb host, we have performed first-principles spin polarized,
semi-relativistic supercell electronic structure calculations within local
spin density approximation (LSDA), employing tight binding linear
muffin-tin orbital method in atomic sphere approximation (TB-LMTO-ASA) [7]
and Von-Barth-Hedin [8] parameterization for the exchange correlation
potential. Calculations were carried out for a single Mo impurity at
substitutional as well as octahedral interstitial site, using cubic
supercells (space group Pm3m) of dimension twice the lattice constant of
fcc Yb, with 32 (33 for interstitial case) atoms and treating Yb-4f as
band electrons.  In order to accommodate the large Mo impurity in the
interstitial site without violating the overlap criterion prescribed by
ASA, the nearest neighbour (nn) Yb atoms had to be relaxed outwards by at
least 5$\%$ of the nn distance. The next nearest neighbour atoms were left
unrelaxed thus keeping the unit cell volume intact.  No relaxation was
necessary for treating the substitutional case.

Figure 4 show the up and down spin local density of states (LDOS) of Mo-d
electron. The results showing large splitting of the two spin sub bands
clearly indicate the presence of high local 4d moments on Mo.  The
calculated moments, summarized in Table I, turn out to be 3.56$\mu_{B}$
for substitutional and 1.12$\mu_{B}$ for the interstitial Mo atoms. For
further confirmation of the rather high moment of Mo, especially at the
interstitial site, we performed additional calculations using spin
polarized full-potential linearized augmented plane wave (FLAPW) method as
implemented in WIEN code [9].  The resulting Mo moments 3.21$\mu_{B}$ for
substitutional and 1.29$\mu_{B}$ for the interstitial sites closely agree
with the LMTO-ASA results. For the interstitial case, calculations with
higher lattice relaxation, 17$\%$ taken from assumption of hard sphere
atomic radii, yielded larger moment $\sim$1.8$\mu_{B}$ (2.1$\mu_{B}$ from
FLAPW method).

Coming back to the experimental results, the Mo magnetic moments can be
estimated from the Curie constants using C = g(S+1)$\mu_B$B(0)/3k$_B$ [5]
where B(0) is the hyperfine field at 0K. For half filled 4d-shell,
neglecting orbital contribution, the measured B(0) consists of a negative
core polarization field, B$_{CP}$ and a positive term (B$_{val}$) arising
from the valence electrons.  The net spin contact hyperfine fields for
many 4d impurities have been found to lie between -200 kG and -280 kG
[10].  Assuming B(0) = -240 kG, the Mo moment for the minority fraction
($\beta(T) <$1) turned out to be 3.96 $\mu_{B}$ which is close to the
value calculated for substitutional site. This supports our earlier
attribution of the $\beta(T) <$1 component to Mo atoms at substitutional
lattice site.  For the majority fraction, the observed $\beta(T) >$1
behaviour implies B(0) to be positive which can arise from a larger
contribution from the valence 5s electrons and much reduced value of
B$_{CP}$. The exact value of B(0) though difficult to calculate, one can
get a reasonable estimate from the $\tau_{N}$ data [11] which came out to
be $\sim$ +19$\pm$(2) T. With this B(0), the Mo moment for majority
fraction was found to be 1.5$\pm 0.3\mu_{B}$, in accordance with the
values calculated for octahedral interstitial site. This gives credence to
our assumption that the component with $\beta(T) >$1 arises from Mo atoms
at octahedral interstitial site.  The above site assignment is also
supported from results reported for Fe in Yb [4]. Here, we like to
emphasize that any uncertainty in B(0) and the consequent spread in the Mo
moments does not influence the main conclusions of this work.

We now examine the stability of Mo magnetic moments in Yb which can be
scaled with the Kondo temperature, T$_{K}$. The later can be derived from
the Curie-Weiss fit of $\beta(T)$ data.  The measured $\beta(T)$
corresponding to substitutional as well as interstitial sites in fcc Yb
yielded T$_{K}$ = 60$\pm$10 K. For Mo in hcp Yb the T$_{K}$ values for the
two sites were found to be 35$\pm$5 K. An estimation of T$_{K}$ could also
be obtained from the spin relaxation rates $\tau_{J}$ extracted from the
$\tau_{N}$ data [11,12].  The T$_{K}$ derived from $\tau_{N}$ data again
turn out to be similar for both substitutional and interstitial sites,
being $\sim$ 55 K for fcc Yb and $\sim$ 20 K for hcp Yb. The $\beta(T)$
and $\tau_{N}$ results indicate rather high moment stability for Mo at
substitutional as well as interstitial lattice sites.

Finally, by analyzing the Kondo temperatures of site specific Mo moments
in Yb we show that magnetism, particularly moment stability, is strongly
influenced by the sign and strength of spin polarization of host
conduction band electrons. Starting from Kondo model, instability of a
magnetic moment is caused by antiferromagnetic exchange interaction
between impurity-d and host conduction electrons. The degree of
instability proportional to T$_{K}$ is governed mainly by the Kondo
resonance width [12] $\Gamma = \pi N(E_{F})V_{kd}^{2}$, where V$_{kd}$ is
the hybridization strength. For divalent Yb with dominantly sp type
conduction electrons, the hybridization strength can be roughly estimated
using the procedure given in Ref [13] and are listed in Table I.  Using
these V$_{kd}$ values the T$_{K} (=\Gamma/k_{B})$ of substitutional Mo
impurity in fcc Yb was found to be $\sim$75 K in close agreement with the
value 60 K measured experimentally.  Further more, the magnetism of
substitutional Mo atom in Yb is consistent with the trend observed in
alkali and alkaline earth metals [11,14] where by the reduction in moment
correlates with the hybridization strength. The above analysis show that
magnetism of substitutional Mo in Yb can be well understood within Kondo
model.  Extrapolating the same physical picture to Mo impurity at
interstitial site, due to stronger hybridization strength (see Table I)
one would expect the moment to be highly unstable with T$_{K} > 400 K$
leading close to nonmagnetic behaviour with $\beta(T)$ = 1.  Instead,
interstitial Mo atoms in Yb  show rather stable moment with Curie-Weiss
type $\beta(T)$ and a low T$_{K} \le$ 60 K.

What is the physical reason for the high moment stability of interstitial
Mo atom in Yb?  To understand this, we look into the host spin
polarization by examining the induced moments at the Yb sites. The results
listed in Table I clearly reveal that sign and strength of host
polarization for the interstitial site qualitatively differ from the
features seen for substitutional case. For substitutional Mo, the small
negative moment at Yb site, mainly arising from sp-band electrons, implies
an antiferromagnetic polarization of host band electrons.  In contrast, Yb
atoms surrounding the interstitial Mo impurity show substantial positive
moment largely due to ferromagnetic polarization of 4f5d band electrons of
Yb. From the results presented above, we believe that this induced
ferromagnetic polarization causing strong interatomic interaction between
Mo-4d and host conduction electrons is mainly responsible for the high
moment stability of interstitial Mo atoms in Yb.  The features of host
polarization found for interstitial Mo in Yb and its influence on T$_{K}$
show striking similarity with the results observed for 3d, 4d impurities
in some d band metals viz.  Pd and PdFe alloys [3,15,16]. As a plausible
mechanism, we suggest that interatomic ferromagnetic interaction between
Mo-4d and host conduction band electrons can compete and successfully
suppress T$_{K}$ arising from antiferromagnetic d-sp exchange interaction
and thereby stabilize the magnetic moment of Mo at the interstitial
lattice site. The above physical picture is consistent with prediction of
recent theoretical calculations where a ferromagnetic interaction between
the impurity and host conduction electrons has been shown to suppress the
Kondo resonance at Fermi energy [17].

To conclude, combining TDPAD experiments with local spin density
calculations, large 4d local moments have been observed for Mo atoms at
substitutional and octahedral interstitial lattice sites of fcc and hcp Yb
metal. While, the magnetism of Mo, for the substitutional site is
consistent with Kondo type antiferromagnetic d-sp exchange interaction, we
find that magnetism and Kondo temperature of interstitial Mo atom is
strongly influenced by ferromagnetic polarization of host Yb-4f5d band
electrons. The results and interpretations presented in this letter
provide an important basis for understanding local magnetism of
interstitial d-impurities in a metallic host. They also yield insight on
the key role of host polarization on the occurrence and stability of local
moments in general.

We thank our Pelletron  staff for their excellent cooperation during the
experiments.

\newpage

\vspace{1.5cm}
\begin{figure}
\caption{}
Spin rotation spectra, R(t)  (left panel) and their Fourier transforms
(right panel) for $^{94}$Mo in fcc and hcp Yb.
\end{figure}

\begin{figure}
\caption{}
Local susceptibility $\beta(T)$ of Mo in fcc (filled symbols) and hcp 
(open symbols) Yb as a function of 1/T. The solid lines correspond to 
fits by Curie-Weiss law: $\beta(T)-1 = C/(T+T_{K})$.
\end{figure}

\begin{figure}
\caption{}
Nuclear relaxation time $\tau_{N}$ as a function of temperature for
$^{94}$Mo in fcc Yb (filled symbols) and hcp Yb (open symbols). The
linear dependence of $\tau_{N}$ with T (solid lines) is indicative of
Korringa like relaxation process (see text).
\end{figure}

\begin{figure}
\caption{}
Local density of states (LDOS) for a Mo impurity in fcc Yb host occupying
substitutional and relaxed (5$\%$) octahedral interstitial sites.
\end{figure}

\begin{table}
\caption{}
Summary of Kondo temperature T$_{K}$, hybridization strength V$_{kd}$  and
calculated (LMTO method) magnetic moments, m$_{0}$ for a Mo impurity at
substitutional and relaxed (5$\%$ and 17$\%$) octahedral interstitial
lattice sites in fcc Yb. m$_{1}$ and m$_{2}$ are the induced moments at
nearest and next nearest Yb atoms. 

\begin{tabular}{|l|c|c|}
                 & {\sl Substitutional} & {\sl Interstitial}     \\ \hline
T$_{K}$(K) & 60     & 60  \\
V$_{kd}$(eV)&0.104   &0.256 \\
m$_{0}(\mu_{B})$ &  3.56    & 1.12(1.77$^{+}$) \\
m$_{1}(\mu_{B})$ & -0.013  & 0.12(0.29$^{+}$)\\
m$_{2}(\mu_{B})$ &  0.008  & 0.0(0.06$^{+}$) \\ 
\end{tabular}
+ correspond to results obtained with 17$\%$ lattice relaxation.
\end{table}


\begin{references}
\bibitem{*} Spin relaxation rate $\tau_{J}^{-1}$ of Mo
can be extracted from $\tau_{J}^{-1} =
2(g_{N}\mu_{N}/\hbar)^{2}B(0)^{2}(S+1)S^{-1}\tau_{N}$ (Ref.5). A rough
estimation of B(0) can be obtained using the facts $\tau_{J}^{-1} =
k_{B}T_{K}$ and $\tau_{N} \propto B(0)^{2}$.
\end{references}
\end{document}